\documentclass[a4paper,12pt]{article}   

\usepackage{amsmath,amsthm,amssymb,cite,enumerate} 

\usepackage[dvips]{color}

\def \BEA { \begin{eqnarray}}
\def \EEA {\end{eqnarray}}
\def \BE {\begin{equation}}
\def \EE {\end{equation}}
\def\d{\mathrm{d}}

\newcommand{\OO}[1] {{O}(r^{-#1})}
\newcommand{\OOp}[1] {{O}(r^{#1})}

\newcommand{\oo}[1] {o(r^{-#1})}
\newcommand{\oop}[1] {o(r^{#1})}

\newcommand{\R}[1] {\rho_{#1}}		

\newcommand{\Om}[1] {\Omega_{#1}}	

\newcommand{\N}[1] {N_{#1}}

\def\a{\alpha}
\def\b{\beta}
\def\g{\gamma}
\def\de{\delta}

\def \mi {\stackrel{i}{m}}
\def \mj {\stackrel{j}{m}}
\def \mk {\stackrel{k}{m}}
\def \mr {\stackrel{r}{m}}
\def \ms {\stackrel{s}{m}}

\def \mz {\stackrel{z}{m}}
\def \mq {\stackrel{q}{m}}
\def \mo {\stackrel{o}{m}}
\def \mD {\stackrel{2}{m}}
\def \mT {\stackrel{3}{m}}
\def \mC {\stackrel{4}{m}}

\def \mio #1 {\mi_{#1}\ ^{  \! \! \! \! 0}} 
\def \mjo #1 {\mj_{#1}\ ^{  \! \! \! \! 0}} 
\def \mko #1 {\mk_{#1}\ ^{  \! \! \! \! 0}} 
\def \mro #1 {\mr_{#1}\ ^{  \! \! \! \! 0}} 
\def \mso #1 {\ms_{#1}\ ^{  \! \! \! \! 0}} 
\def \mpo #1 {\mp_{#1}\ ^{  \! \! \! \! 0}} 
\def \mzo #1 {\mz_{#1}\ ^{  \! \! \! \! 0}} 
\def \mqo #1 {\mq_{#1}\ ^{  \! \! \! \! 0}} 
\def \moo #1 {\mo_{#1}\ ^{  \! \! \! \! 0}} 
\def \mDo #1 {\mD_{#1}\ ^{  \! \! \! \! 0}} 
\def \mTo #1 {\mT_{#1}\ ^{  \! \! \! \! 0}} 
\def \mCo #1 {\mC_{#1}\ ^{  \! \! \! \! 0}} 

\def \miJ #1 {\mi_{#1}\ ^{  \! \! \! \! (1)}} 
\def \mjJ #1 {\mj_{#1}\ ^{  \! \! \! \! (1)}} 
\def \mkJ #1 {\mk_{#1}\ ^{  \! \! \! \! (1)}} 
\def \mrJ #1 {\mr_{#1}\ ^{  \! \! \! \! (1)}}

\def \bl {\mbox{\boldmath{$\ell$}}}

\def \bn {\mbox{\boldmath{$n$}}}

\def \hbm #1 {\mbox{\boldmath{$\hat m^{(#1)}$}}}
\def \bm {\mbox{\boldmath{$m$}}}

\newcommand{\be}{\begin{equation}}
\newcommand{\ee}{\end{equation}}
\newcommand{\beqn}{\begin{eqnarray}}
\newcommand{\eeqn}{\end{eqnarray}}
\newcommand{\pa}{\partial}
\newcommand{\ba}{\begin{array}}
\newcommand{\ea}{\end{array}}

\def \bbm {\mbox{\boldmath{$m$}}}

\def \BEAH {\begin{eqnarray*}}
\def \EEAH {\end{eqnarray*}}
\def \BEA {\begin{eqnarray}}
\def \EEA {\end{eqnarray}}
\def \BDM {\begin{displaymath}}
\def \EDM {\end{displaymath}}

\newcommand{\M}[3] {{\stackrel{#1}{M}}_{{#2}{#3}}}
\newcommand{\m}[3] {{\stackrel{\hspace{.3cm}#1}{m}}_{\!{#2}{#3}}\,}

\begin{document}

\title{Asymptotic behaviour of Maxwell fields in higher dimensions}

\author{Marcello Ortaggio\thanks{ortaggio(at)math(dot)cas(dot)cz} \\
Institute of Mathematics, Academy of Sciences of the Czech Republic \\ \v Zitn\' a 25, 115 67 Prague 1, Czech Republic}
\date{\today}

\maketitle

\begin{abstract}

We study the fall-off behaviour of test electromagnetic fields in higher dimensions as one approaches infinity along a congruence of ``expanding'' null geodesics. The considered backgrounds are Einstein spacetimes including, in particular, (asymptotically) flat and (anti-)de~Sitter spacetimes. Various possible boundary conditions result in different characteristic fall-offs, in which the leading component can be of any algebraic type (N, II or G). In particular, the peeling-off of radiative fields $F=Nr^{1-n/2}+Gr^{-n/2}+\ldots$ differs from the standard four-dimensional one (instead it qualitatively resembles the recently determined behaviour of the Weyl tensor in higher dimensions). General $p$-form fields are also briefly discussed. In even $n$ dimensions, the special case $p=n/2$ displays unique properties and peels off in the ``standard way'' as $F=Nr^{1-n/2}+IIr^{-n/2}+\ldots$. A few explicit examples are mentioned.

\end{abstract}



\section{Introduction}

Formal analogies between the gravitational and the electromagnetic field have proven useful in developing exact approaches in general relativity, particularly in the study of gravitational radiation and asymptotic properties near null infinity. A well-known result is the peeling-off property. For the gravitational field, this means that the Weyl tensor in a frame parallelly transported along a congruence of ``outgoing'' null geodesics $\bl$ in an asymptotically flat spacetime decays as \cite{Sachs61,BBM,Sachs62,NP,Penrose63,Penrose65prs}
\be
 C_{abcd}=\frac{N_{abcd}}{r}+\frac{III_{abcd}}{r^2}+\frac{II_{abcd}}{r^3}+\frac{I_{abcd}}{r^4}+\frac{G_{abcd}}{r^5}+\ldots \qquad (n=4) ,
	\label{Weyl4D}
\ee
where $r$ is an affine parameter along $\bl$, and the increasing powers of $1/r$ describe terms of increasing boost-weight (b.w.) w.r.t. $\bl$ (i.e., decreasing alignment \cite{Milsonetal05} -- the $G_{abcd}$ term is not aligned and is thus ``generic''). Similarly, for the Maxwell field one has (both for test fields in flat space \cite{GK,JanNew65,Penrose65prs} and for the full Einstein-Maxwell theory \cite{Penrose65prs})
\be
 F_{ab}=\frac{N_{ab}}{r}+\frac{II_{ab}}{r^2}+\frac{G_{ab}}{r^3}+\ldots \qquad (n=4) .
	\label{Maxw4D}
\ee

In order to arrive at \eqref{Weyl4D} and \eqref{Maxw4D}, one needs to assume suitable boundary conditions on components of maximal b.w. (in Newman-Penrose notation, $\Psi_0=\OO{5}$ and $\Phi_0=\OO{3}$, respectively), along with sufficient smoothness properties. The decay rate of the remaining field components then automatically follows. Both in \eqref{Weyl4D} and \eqref{Maxw4D} the algebraic type becomes less and less special as one recedes from infinity, and the leading $1/r$ term represents a (``null'') radiative field, which can be related to the flux of radiated energy 
and to the gravitational and electromagnetic news functions \cite{BBM,Sachs62,NewUnt62,Penrose63,ExtNewPen69,vanderBurg69}. Peeling of the fields occurs also in asymptotically (A)dS spacetimes \cite{Penrose65prs}. 

Recently, it has been proven \cite{GodRea12} that in $n>4$ dimensions the peeling behaviour of the Weyl tensor of asymptotically flat spacetimes is qualitatively different from \eqref{Weyl4D}. Furthermore, by dropping the requirement of asymptotic flatness, various possible choices of boundary conditions at infinity (involving also non-maximal b.w. components) result in a rich pattern of different asymptotic decays, also depending non-trivially on the presence of a cosmological constant (as opposed to the case $n=4$) \cite{OrtPra14}. One could thus expect that the behaviour of the Maxwell field $F_{ab}$ in higher dimensions also differs from the 4D result \eqref{Maxw4D}. In this paper we find that this is indeed the case. More specifically, we determine the possible fall-off of the electromagnetic field at null infinity and how it depends on the chosen boundary conditions. Several analogies with the results of \cite{GodRea12,OrtPra14} for the gravitational field will be found.
In particular, when the cosmological constant vanishes and with the choice of boundary conditions $F_{0i}=\OO{\frac{n}{2}}$ and $F_{ij}=\oo{2}$,\footnote{$F_{0i}$ and $F_{ij}$ are frame components of b.w. $+1$ and 0, respectively, defined in section~\ref{subsec_notat}} in more than four dimensions the behaviour \eqref{Maxw4D} is replaced by (cf.~\eqref{peeling_gen})
\be
 F_{ab}=\frac{N_{ab}}{r^{\frac{n}{2}-1}}+\frac{G_{ab}}{r^\frac{n}{2}}+\ldots \qquad (n>4) . 
	\label{Maxw_higher}
\ee
The leading term is null and has a fall-off rate characteristic of radiation as in 4D, but already the first subleading term is algebraically general and differs from \eqref{Maxw4D}. Furthermore,  it turns out that if $\bl$ is aligned with $F_{ab}$ then the radiative term necessarily vanishes when $n>4$, in analogy with the gravitational case \cite{GodRea12,OrtPraPra09b,OrtPra14} -- more generally this occurs whenever $F_{0i}=\oo{\frac{n}{2}}$.

Similarly as in \cite{OrtPra14}, for $n>4$ various other choices of boundary conditions for $F_{ab}$ are possible, and the corresponding asymptotic fall-offs (also including a cosmological constant) will be detailed in the paper (sections~\ref{R=0} and \ref{sec_R}). However, in addition to the standard 2-form Maxwell field $F_{ab}$, in higher dimensional theories general $p$-form fields $F_{a_1\ldots a_p}$ may also play an important role (see, e.g., \cite{HenTei86}). For a generic $p$, their asymptotic behaviour turns out to be similar to that of the case $p=2$ (including \eqref{Maxw_higher}), as briefly discussed in section~\ref{sec_p}. In even $n$ dimensions, the case $p=\frac{n}{2}$ is however peculiar, since it is the only case in which Maxwell's equations are conformally invariant and admit self-dual solutions (for odd $p$). Correspondingly, we find that the asymptotic behaviour of $\frac{n}{2}$-form fields is also exceptional, and in fact qualitatively similar to the case $n=4$, $p=2$ (eq.~\eqref{Maxw4D}), cf.~\eqref{peeling_n/2}.\footnote{In \cite{Durkeeetal10} it was pointed out that type N Maxwell fields with $p=\frac{n}{2}$ possess special ``optical'' properties (see also \cite{HugMas88} for related earlier results). This is indeed related to the special peeling we find, see further comments in sections~\ref{subsubsec_a<-2_subcase} and \ref{subsec_n/2}.}

We emphasize that in this paper we restrict to {\em test} fields satisfying the source-free Maxwell equations in the background of a certain class of Einstein spacetimes \cite{OrtPra14} (including, in particular, spaces of constant curvature -- details in section~\ref{sec_notat}). We will generally study the $r$-dependence (for $r\to\infty$) of the leading field components only, under the assumption that this is {\em power-like}  (some comments on certain subleading terms will however be necessary to arrive, e.g., at \eqref{Maxw_higher}). It will thus not be necessary to assume that the field admits a power-series expansion in $1/r$ (however, the existence of full Einstein-Maxell solutions with that property is proven in \cite{ChoChrLoi06}, in the case of even dimensions).
More technical assumptions will be explained in section~\ref{sec_notat}, where certain results of \cite{OrtPra14} needed in this work will also be summarized. In sections \ref{R=0} and \ref{sec_R} we present the full asymptotic behaviour of $F_{ab}$ for Ricci flat and proper Einstein spacetimes, respectively, while most of the related technical details (including the full Maxwell equations in a ``null'' frame) are relegated to appendix~\ref{app_setup} (appendix~\ref{app_Ricci} simply contains the definition of the Ricci rotation coefficients). For the case $p=2$, in appendix~\ref{app_transv_R=0} we give the ``transverse'' (i.e., non-radial) Maxwell equations in the various permitted cases (for $\tilde R=0$ only, since when $\tilde R\neq 0$ many of those turn out to be identically satisfied at the leading order). They are not needed to obtain asymptotic properties of the field, yet they provide explicit relations between leading components of different b.w., and can be viewed as a preliminary step in the direction of a systematic analysis of asymptotic solutions of the Maxwell equations.  Appendix~\ref{app_4D} summarizes main results for the four-dimensional case.

\section{Preliminaries}

\label{sec_notat}

\subsection{Assumptions and notation}

\label{subsec_notat}

We consider test Maxwell fields in the background of an $n$-dimensional Einstein spacetime ($R_{ab}=\frac{R}{n}g_{ab}$), with $n\ge4$. We restrict to Einstein spacetimes with suitable properties at null infinity, which we formulate in terms of a congruence of outgoing null geodesics and of the Weyl tensor. Namely, let $\bl=\pa_r$ be a geodesic null vector field parametrized by an affine parameter $r$. First, we assume that the optical matrix $\R{ij}$ of $\bl$ (defined below in \eqref{rho}) is {\em asymptotically non-singular and expanding}, i.e., the leading term of $\R{ij}$ (for $r\to\infty$) is a matrix with non-zero determinant and non-zero trace. This essentially means that, close to null infinity, $\bl$ expands in all spacelike directions at the same speed, which is compatible, in particular, with asymptotically flat spacetimes \cite{TanKinShi11,GodRea12} but also holds more generally \cite{OrtPra14} (see \cite{NP} for a related discussion in four dimensions further motivating this assumption). Next, we assume that in a frame parallelly transported along $\bl$ the b.w. $+2$ components of the Weyl tensor fall off ``fast enough'' along $\bl$, namely 
\be
	\Om{ij}=\oo{3} .
	\label{Omega}
\ee	
This is a rather weak restriction (e.g., in 4D the spacetimes considered in \cite{NP} obey the stronger condition $\Om{ij}=\OO{5}$) and corresponds to the class of Einstein spacetimes recently studied in \cite{OrtPra14}.\footnote{To be precise, only leading terms of Weyl components with a power-like behaviour were considered in \cite{OrtPra14}, and in fact the weaker condition $\Om{ij}=\oo{2}$ was assumed. We could similarly relax \eqref{Omega} with no change in the results of the present paper as far as the $r$-dependence of the Maxwell field is concerned. However, the slightly stronger restriction \eqref{Omega} permits us to determine both the leading and subleading terms in the key equation \eqref{L_2}, which in turn will enable us to obtain explicitly additional ``constraints'' on the Maxwell field components (for example, those given in point~2. of section~\ref{subsec_bw0}).\label{footn_omega}} 
It contains, in particular, asymptotically flat spacetimes \cite{GodRea12,OrtPra14}, algebraically special spacetimes for which $\bl$ is a WAND (i.e., a Weyl aligned null direction -- in this case $\Om{ij}=0$ identically and \eqref{Omega} is satisfied trivially), and of course Minkowski and (anti-)de~Sitter spaces when the Weyl tensor vanishes identically. Under the above conditions one is able to determine how the various Maxwell tensor components fall off as $r\to\infty$, as shown in what follows.

Throughout the paper we employ a frame adapted to $\bl$ consisting of two null vectors $\bm_{(0)}=\bl$ and $\bm_{(1)}=\bn$, and $n-2$ orthonormal spacelike vectors $\bm_{(i)}$,  where $i, j, \ldots=2,\ldots,n-1$, such that $\bl \cdot \bn =1$, $\bm_{(i)} \cdot \bm_{(j)} = \delta_{ij}$ and  $\bl \cdot \bm_{(i)} = 0 = \bn \cdot \bm_{(i)}$. In terms of these one can define the frame Weyl and Ricci tensor components and the Ricci rotation coefficients, for which we use the notation summarized in the recent review \cite{OrtPraPra13rev}. In particular, the Weyl components \eqref{Omega} are given by
\be
 \Om{ij}\equiv C_{0i0j}=C_{abcd}\ell^a m_{(i)}^b\ell^c m_{(j)}^d ,
\ee
and the $(n-2)\times(n-2)$ optical matrix $\R{ij}$ associated with $\bl$ is defined by
\be
 \R{ij} = \ell_{a;b}m_{(i)}^a m_{(j)}^b  , \label{rho}
\ee
whose trace gives the expansion scalar 
\be
 \R{}=\R{ii}.
\ee
The remaining Ricci rotation coefficients are defined analogously \cite{OrtPraPra13rev} (see appendix~\ref{app_Ricci}). Similarly, one can define the components of the Maxwell field $F_{ab}$ in the same frame, namely $F_{0i}=F_{ab}\ell^a m_{(i)}^b$ (b.w. $+1$), $F_{01}=F_{ab}\ell^a n^b$ and $F_{ij}=F_{ab}m_{(i)}^a m_{(j)}^b$ (b.w. 0), and $F_{1i}=F_{ab}n^a m_{(i)}^b$ (b.w. $-1$).

Taking $r$ as one of the coordinates we can write the frame derivative operators as
\be
 D=\pa_r , \qquad \Delta=U\pa_r+X^A\pa_A , \qquad \delta_i=\omega_i\pa_r+\xi^A_i\pa_A ,
\label{der_oper}
\ee
where $\pa_A=\pa/\pa x^A$ and the $x^A$ represent any set of ($n-1$) scalar functions such that $(r,x^A)$ is a well-behaved coordinate system.

\subsection{Asymptotic $r$-dependence of Ricci rotation coefficients and derivative operators}

\label{subsec_r_Ricci}

In the following sections we will study the Maxwell equations, evaluated in the frame defined above, in order to fix the $r$-dependence of the Maxwell tensor. To that end, one first needs to know how the Ricci rotation coefficients and the derivative operators depend on $r$ in the considered spacetimes, as we now summarize. 

The condition that $\bl$ be geodetic and affinely parametrized is expressed by
\be
	\kappa_i=0=L_{10} .
	\label{l_geod}
\ee
Since the frame vectors are taken to be parallelly transported along $\bl$, one also has
\be 
	\M{i}{j}{0}=0, \qquad N_{i0}=0 .
	\label{parall_frame}
\ee

Apart from the above trivial ones, the $r$-dependence of all Ricci rotation coefficients and derivative operators has been obtained  in \cite{OrtPra14}.
Several quantities are not affected by the presence of a cosmological constant at the leading order and are given by \cite{OrtPra14} 
\beqn
 & & \R{ij}=\frac{\delta_{ij}}{r}+\frac{b_{ij}}{r^2}+\ldots , \label{L_2} \\
 & & L_{1i}=\frac{l_{1i}}{r}+\ldots , \qquad L_{i1}=\frac{l_{i1}}{r}+\ldots , \qquad \M{i}{j}{k}=\frac{\m{i}{j}{k}}{r}+\ldots , \qquad \M{i}{j}{1}=\m{i}{j}{1}+\ldots , \\
 & & \xi^A_i=\frac{\xi^{A0}_i}{r}+\ldots, \qquad  \omega_i=-l_{1i}+\ldots , \qquad X^A=X^{A0}+\ldots ,
\eeqn
where $b_{ij}$ is independent of $r$, the ellipsis denote generically subleading terms (of unspecified lower order), and lowercase symbols or a superscript $^0$ denote quantities independent of $r$, which at this stage are arbitrary integration functions (of the corresponding radial Ricci identities or commutators). 
Using null rotations about $\bl$ one could choose a parallelly transported frame such that, e.g., $l_{1i}=0$ or $l_{i1}=0$ -- this may simplify certain expressions in the following, but for the sake of generality we will keep our frame unspecified. Note also that if $\bl$ is twistfree then $b_{[ij]}=0$. 

The behaviour of the remaining Ricci rotation coefficients and derivative operators depends on the presence of a cosmological constant, and is given below separately for Ricci-flat and for proper Einstein spacetimes. In order to have more compact formulas it is convenient to define the rescaled Ricci scalar 
\be 
 \tilde R=\frac{R}{n(n-1)}. 
\label{R_rescaled}
\ee

\subsubsection{Remaining quantities for $R=0$}

\label{Ricci_subsubsec_R=0}

For $R=0$ (i.e., $\tilde R=0$, recall~\eqref{R_rescaled}) one has \cite{OrtPra14}

\beqn
 & & \N{ij}=\frac{n_{ij}}{r}+\ldots , \qquad L_{11}=l_{11}+\ldots , \qquad N_{i1}=n_{i1}+\ldots , \\
 & & U=-l_{11}r+\ldots .
\eeqn

\subsubsection{Remaining quantities for $R\neq 0$}

\label{Ricci_subsubsec_R}

In the presence of a cosmological constant one finds instead \cite{OrtPra14}

\beqn
 & & \N{ij}=-\frac{\tilde R}{2}\delta_{ij}r+\frac{\tilde R}{2}b_{ij}\ldots , \qquad L_{11}=\tilde Rr+\ldots , \qquad N_{i1}=\frac{\tilde R}{2}l_{i1}r+\ldots , \label{Ricci_R} \\
 & & U=-\frac{\tilde R}{2}r^2+\ldots . \label{U_R} 
\eeqn
(Note that for $\N{ij}$ we have also given the subleading term, obtained in \cite{OrtPra14} under the assumption it is also power-like. While this is not necessary in order to determine the $r$-dependence of the leading order components of the Maxwell field, it will enable us to write down some constraints coming from the non-radial Maxwell equations -- cf. also footnote~\ref{footn_omega}).

\subsection{Maxwell's equations}

In the frame defined in section~\ref{subsec_notat}, with the definitions \eqref{der_oper} and with the conditions \eqref{l_geod} and \eqref{parall_frame}, the empty-space Maxwell equations $F^a_{\ \; b;a}=0$ and $F_{[ab;c]}=0$ take the form \eqref{A1}--\eqref{D2} given in appendix~\ref{app_setup}. Now, we assume that for $r\to\infty$
\be
 F_{0i}=F_{0i}^{(\a)}r^\a+\ldots , \qquad F_{01}=F_{01}^{(\b)}r^\b+\ldots , \qquad F_{ij}=F_{ij}^{(\g)}r^\g+\ldots , \qquad F_{1i}=F_{1i}^{(\de)}r^\de+\ldots , \label{leading}
\ee
where $F_{0i}^{(\a)}$, $F_{01}^{(\b)}$, $F_{ij}^{(\g)}$ and $F_{1i}^{(\de)}$ do not depend on $r$. We will also assume that if a generic component $f$ behaves as $f=\OO{\zeta}$, then $\pa_rf=\OO{\zeta-1}$ and $\pa_Af=\OO{\zeta}$. We are interested in determining the values of $\a$, $\b$, $\g$ and $\de$ (a priori not restricted to be integers or semi-integers) compatible with Maxwell's equations. From now on it will be understood that the results of section~\ref{subsec_r_Ricci} will be employed, when necessary.

The study of the Maxwell components of  b.w. 0 shows that several possible cases appear, essentially depending on $\a$, $\b$ and $\g$ being greater, equal, or smaller than certain ``critical'' values. This is related to a freedom in choosing boundary conditions at infinity. Then, by looking at components of b.w. $-1$, also the permitted values of $\de$ becomes sensitive to the boundary conditions one chooses. All these technical details are relegated to appendix~\ref{app_setup}. In the following two sections we present the results obtained in the case of a zero and a non-zero cosmological constant.

\section{Full asymptotic behaviour for $R=0$} 

\label{R=0}

Using the intermediate steps described in appendix~\ref{app_setup}, one arrives at the following results, which is necessary to give  separately for $n\ge6$ and  $n=5$.

\subsection{Case $n\ge6$}

\label{subsec_nge6}

\subsubsection{Fall-off of $F_{0i}$ with $\a\ge-2$}

\label{subsubsec_a>=-2}

If $\a\ge-2$ then 
	\beqn
		& & F_{0i}=F_{0i}^{(\a)}r^\a+\ldots , \label{F0i_case1} \\
		& & F_{01}=F_{01}^{(\a)}r^\a+\ldots , \qquad F_{ij}=F_{ij}^{(\a)}r^\a+\ldots , \\
		& & F_{1i}=F_{1i}^{(\a)}r^\a+\ldots , \label{F1i_case1}
	\eeqn
	where $F_{01}^{(\a)}$ and $F_{ij}^{(\a)}$ are given by \eqref{F01_gener} and \eqref{Fij_gener}, $F_{1i}^{(\a)}$ is given (except for $\a=1-\frac{n}{2}$) using \eqref{-C1+A2} by
	\beqn
		(n-2+2\a)F_{1i}^{(\a)}=F_{0j}^{(\a)}(n_{ij}+n_{ji})-F_{0i}^{(\a)}n_{jj}+\xi^{A0}_jF_{ij,A}^{(\a)}+F_{ij}^{(\a)}(-\a l_{1j}+\m{j}{k}{k})+F_{jk}^{(\a)}\m{j}{i}{k} \nonumber \\
								{}-\xi^{A0}_iF_{01,A}^{(\a)}+\a F_{01}^{(\a)}l_{1i} , \label{-C1+A2_gener}
	\eeqn	
	and conditions~\eqref{+C1+A2_gener}--\eqref{D2_case1} hold.

Clearly in the present case (eqs.~\eqref{F0i_case1}--\eqref{F1i_case1}) the electromagnetic field does not peel. For $\a\ge-\frac{3}{2}$ the same results apply also for $n=5$ (cf. section~\ref{subsubsec_n=5_a>=-3/2}). For example, one can verify that a uniform magnetic field in the background of asymptotically flat black holes \cite{AliFro04} or black rings \cite{OrtPra06} has such a behaviour (with $\a=0$).

\subsubsection{Fall-off of $F_{0i}$ with $\a<-2$: generic case}

\label{subsubsec_a<-2}

	If $\a<-2$ we generically have 
	\beqn
		& & F_{0i}=F_{0i}^{(\a)}r^\a+\ldots , \label{+1_gamma=-2} \\
		& & F_{01}=\oo{2} , \qquad F_{ij}=\frac{F_{ij}^{(-2)}}{r^2}+\ldots , \label{0_gamma=-2} \\
		& & F_{1i}=\frac{F_{1i}^{(-2)}}{r^2}+\ldots \label{F1i_case2} . \label{-1_gamma=-2}
	\eeqn

Recall that additional conditions hold that determine the leading term of $F_{01}$ -- cf. point 2. of section~\ref{subsec_bw0}. The above behaviour includes the special case when $\bl$ is a an aligned null direction of the Maxwell field, i.e., $F_{0i}=0$ (in the formal limit $\a\to-\infty$). The leading term falls as $1/r^2$ and is of type II. In cases with $F_{1i}^{(-2)}=0$ the leading field components are purely magnetic. Explicit examples can be obtained as a ``linearized'' Maxwell field limit of certain full Einstein-Maxwell solutions given in \cite{OrtPodZof08} for even $n$.

Further conditions are given by eq.~\eqref{B2_case2}--\eqref{constr_a<-3_gamma=-2}.

\subsubsection{Fall-off of $F_{0i}$ with $\a<-2$: subcase $F_{ij}^{(-2)}=0$}

\label{subsubsec_a<-2_subcase}

So far we have described the generic case $\a<-2$ with $F_{ij}^{(-2)}\neq0$. However, if the magnetic term $F_{ij}^{(-2)}$ vanishes we have the following subcases. 

	\begin{enumerate}[(a)]

			\item\label{n>5_gen} For $1-\frac{n}{2}\le\a<-2$ we have the same results as in section~\ref{subsubsec_a>=-2} above (i.e., eqs.~\eqref{F0i_case1}--\eqref{-C1+A2_gener} with \eqref{+C1+A2_gener}--\eqref{D2_case1}). Note that this subcase does not exist for $n=6$.
						
			\item\label{n>5_rad} For $-\frac{n}{2}\le\a<1-\frac{n}{2}$ we have 
				\beqn
					& & F_{0i}=F_{0i}^{(\a)}r^\a+\ldots , \label{+1_rad} \\
					& & F_{01}=F_{01}^{(\a)}r^\a+\ldots , \qquad F_{ij}=F_{ij}^{(\a)}r^\a+\ldots , \label{0_rad} \\
					& & F_{1i}=\frac{F_{1i}^{(1-\frac{n}{2})}}{r^{\frac{n}{2}-1}}+\ldots , \label{-1_rad} 
			  \eeqn
					with \eqref{F01_gener}, \eqref{Fij_gener}. If $-\frac{n}{2}<\a<1-\frac{n}{2}$ then the constraints \eqref{+C1+A2_gener}--\eqref{D2_case1} hold, whereas for the special (semi-)integer value $\a=-\frac{n}{2}$ one has (from \eqref{+C1+A2})
					
					\be 
					  (n-4)F_{1i}^{(1-\frac{n}{2})}=-2\left[X^{A0}F_{0i,A}^{(-\frac{n}{2})}+\left(\frac{n}{2}-1\right)F_{0i}^{(-\frac{n}{2})}l_{11}+F_{0j}^{(-\frac{n}{2})}\m{j}{i}{1}\right] \qquad\left(\a=-\frac{n}{2}\right) , \label{+C1+A2_rad}
					\ee
					with \eqref{B1_case2_rad}--\eqref{B2_case2_rad}.

At infinity, the leading term of the Maxwell components \eqref{+1_rad}--\eqref{-1_rad} falls off as $1/r^{\frac{n}{2}-1}$ and is of type N. This is characteristic of radiative fields ($T_{11}\propto F_{1i}F_{1i}\sim 1/r^{n-2}$, i.e., the energy flux along $\bl$, can be directly related to the energy loss, at least in the case of asymptotically flat spacetimes -- cf.~\cite{Penrose63,ExtNewPen69,vanderBurg69} for $n=4$). Note that $\bl$ cannot be aligned with $F_{ab}$ if radiation is present (since $\a\ge-\frac{n}{2}$), as opposed to the well-known four-dimensional case. Furthermore, in the case $\a=-\frac{n}{2}$, if one assumes that $F_{1i}$ has a power-like behaviour also at the subleading order, from \eqref{-C1+A2} one finds $F_{1i}=F_{1i}^{(1-\frac{n}{2})}{r^{1-\frac{n}{2}}}+\OO{n/2}$. This implies the peeling-off behaviour
		\be
		 F_{ab}=\frac{N_{ab}}{r^{\frac{n}{2}-1}}+\frac{G_{ab}}{r^{\frac{n}{2}}}+\ldots \qquad\left(\a=-\frac{n}{2}\right) .
		\label{peeling_gen}
		\ee 
		(Without assuming $\a=-\frac{n}{2}$ the subleading term would be of order $\OOp{\a}$.)
		We observe that the subleading term is algebraically general, which is qualitatively different from the 4D case~\eqref{Maxw4D}. This resembles the behaviour of the Weyl tensor of higher dimensional asymptotically flat spacetimes found in \cite{GodRea12} (see also \cite{OrtPra14}). See section~\ref{R=0_5D_special} for a possible different peeling-off in five dimensions.

			\item\label{n>5_subrad} For $2-n\le\a<-\frac{n}{2}$ we have	the same results as in section~\ref{subsubsec_a>=-2} above.	
			
			\item\label{n>5_Coulomb} For $\a<2-n$ we have 
				\beqn
					& & F_{0i}=F_{0i}^{(\a)}r^\a+\ldots , \label{+1_n>5_Coulomb} \\
					& & F_{01}=\frac{F_{01}^{(2-n)}}{r^{n-2}}+\ldots , \qquad F_{ij}=\oop{2-n} , \\
					& & F_{1i}=\frac{F_{1i}^{(2-n)}}{r^{n-2}}+\ldots , \label{-1_n>5_Coulomb}
			  \eeqn
			with (from \eqref{-C1+A2})
				\be
					F_{1i}^{(2-n)}=\frac{1}{n-2}\xi^{A0}_iF_{01,A}^{(2-n)}+F_{01}^{(2-n)}l_{1i} .
					\label{-C1+A2_case2_a<2-n}
				\ee
				
Recall that additional conditions hold that determine the leading term of $F_{ij}$ -- cf. point 2.(B) of section~\ref{subsec_bw0}. 

The leading term is of type II and falls off as $1/r^{n-2}$. It becomes purely electric in the case $F_{1i}^{(2-n)}=0$. This behaviour describes, in particular, the Coulomb field of a weakly charged asymptotically flat black hole \cite{AliFro04,Aliev06prd} or black ring\footnote{To be precise, the background black ring spacetime \cite{EmpRea02prl} has $n=5$ and should thus be considered in section~\ref{subsec_n=5} below. However, as it turns out, the asymptotic behaviour described here applies also for $n=5$ (cf. section~\ref{subsubsec_R=0_n=5_a<-5/2}), thus there is no need to mention the black ring case separately.} \cite{OrtPra06} (where $F_{ij}=F_{ij}^{(1-n)}r^{1-n}+\ldots$ represents magnetic dipoles induced by rotation). One further has eqs.~\eqref{case2B_a<1-n_first}--\eqref{case2B_a<1-n_last}.

			In the special subcase $F_{ij}^{(-2)}=0=F_{01}^{(2-n)}$ with $\a<2-n$ the same results as in section~\ref{subsubsec_a>=-2} again apply (in particular, for  $n=5$ and $\a=-4$ this is the case of the weak-field limit of the 5D dipole black rings of \cite{Emparan04}). This means that if the Maxwell tensor is not of type G and $\bl$ is taken to be an aligned null direction, then $F_{ab}=0$ identically in this special case. Therefore, fields more special than type G and aligned with $\bl$ require either $F_{ij}^{(-2)}\neq0$ or $F_{01}^{(2-n)}\neq0$ (or both), so that type N fields for which $\bl$ is doubly aligned are not permitted here.\footnote{In fact, type N fields singly aligned with $\bl$ are not permitted either (since a type N Maxwell field can have only one aligned null direction -- which must thus be doubly aligned by definition of type N).} The latter property is due to the fact that an expanding aligned null direction of a type N Maxwell field is necessarily shearing when $n>4$ \cite{Ortaggio07} (see \cite{Durkeeetal10} for extensions of this result) while our $\bl$ is not (at least at the leading order), by construction (eq.~\eqref{L_2}).

	\end{enumerate}

\subsection{Case $n=5$}

\label{subsec_n=5}

\subsubsection{Fall-off of $F_{0i}$ with $\a\ge-\frac{3}{2}$}

\label{subsubsec_n=5_a>=-3/2}

For $n=5$ with $\a\ge-\frac{3}{2}$ the same results as for $n\ge6$ with $\a\ge-2$ still apply, i.e., those of section~\ref{subsubsec_a>=-2}. This includes the already mentioned uniform magnetic fields.

\subsubsection{Fall-off of $F_{0i}$ with $-2\le\a<-\frac{3}{2}$}

\label{subsubsec_5D_rad1}

If $-2\le\a<-\frac{3}{2}$ we obtain
	\beqn
			& & F_{0i}=F_{0i}^{(\a)}r^\a+\ldots , \label{+1_5D_rad1} \\
			& & F_{01}=F_{01}^{(\a)}r^\a+\ldots , \qquad F_{ij}=F_{ij}^{(\a)}r^\a+\ldots ,  \qquad\qquad (n=5) \label{0_5D_rad1} \\
			& & F_{1i}=\frac{F_{1i}^{(-\frac{3}{2})}}{r^{\frac{3}{2}}}+\ldots , \label{-1_5D_rad1} 
	\eeqn
with \eqref{F01_gener}, \eqref{Fij_gener}. This is the same radiative behavior as the one described by \eqref{+1_rad}--\eqref{-1_rad} when $n\ge6$, however here we have $\a\ge-2$ and thus always $\a>-\frac{n}{2}=-\frac{5}{2}$. From \eqref{+C1+A2}--\eqref{D2} we thus obtain, respectively, \eqref{+C1+A2_gener}--\eqref{D2_case1}.

\subsubsection{Fall-off of $F_{0i}$ with $-\frac{5}{2}\le\a<-2$}

\label{R=0_5D_special}

If $-\frac{5}{2}\le\a<-2$ we obtain
	\beqn
			& & F_{0i}=F_{0i}^{(\a)}r^\a+\ldots , \\
			& & F_{01}=F_{01}^{(\a)}r^\a+\ldots , \qquad F_{ij}=\frac{F_{ij}^{(-2)}}{r^2}+\ldots , \qquad\qquad (n=5) \\
			& & F_{1i}=\frac{F_{1i}^{(-\frac{3}{2})}}{r^{\frac{3}{2}}}+\ldots ,
	\eeqn
with \eqref{F01_gener}.	
We observe that there is no such a behavior when $n\ge6$ (cf. \eqref{+1_gamma=-2}--\eqref{-1_gamma=-2}).	If $-\frac{5}{2}<\a<-2$ then \eqref{+C1+A2_gener} holds, whereas for $\a=-\frac{5}{2}$ one has \eqref{+C1+A2_rad} (with $n=5$), which determines $F_{1i}^{(-\frac{3}{2})}$. In both cases we further have \eqref{B2_case2} and \eqref{D2_case2} (from \eqref{B2}, \eqref{D2}). Recall also the comments in point 2. of section~\ref{subsec_bw0}. Depending on the value of $\a$, \eqref{B1} gives \eqref{B1_case1} for $-\frac{5}{2}<\a<-2$, and \eqref{B1_case2_rad} for $\a=-\frac{5}{2}$.

If one assumes that $F_{1i}$ has a power-like behaviour also at the subleading order, from \eqref{-C1+A2} one finds $F_{1i}=F_{1i}^{(-\frac{3}{2})}r^{-\frac{3}{2}}+\OO{2}$ (with \eqref{-C1+A2_gamma=-2}). This implies the peeling-off behaviour (for any $-\frac{5}{2}\le\a<-2$)
		\be
		 F_{ab}=\frac{N_{ab}}{r^{\frac{3}{2}}}+\frac{II_{ab}}{r^2}+\ldots \qquad (n=5) ,
		\label{peeling_5D}
		\ee 
		which at the subleading order differs from the general $n\ge6$ behaviour \eqref{peeling_gen}, due to the magnetic term $F_{ij}^{(-2)}$. It is however a question whether full 5D Einstein-Maxwell solutions exist with $F_{ij}^{(-2)}\neq0$ -- results of \cite{OrtPodZof08} show that this is not the case, at least when $\bl$ is shearfree and twistfree and $F_{ab}$ is aligned with $\bl$.

A special subcase occurs when $F_{ij}^{(-2)}=0$. For $-\frac{5}{2}<\a<-2$ this leads to all the same results as in section~\ref{subsubsec_5D_rad1}, while for $\a=-\frac{5}{2}$ one still has the asymptotic behaviour \eqref{+1_5D_rad1}--\eqref{-1_5D_rad1} but with the conditions \eqref{+C1+A2_rad} and \eqref{B1_case2_rad}--\eqref{B2_case2_rad}.

\subsubsection{Fall-off of $F_{0i}$ with $\a<-\frac{5}{2}$}

\label{subsubsec_R=0_n=5_a<-5/2}

When $\a<-\frac{5}{2}$ one obtains the same results as for $n\ge 6$ with $\a<-2$, i.e., the fall-off \eqref{+1_gamma=-2}--\eqref{-1_gamma=-2} with eqs.~\eqref{B2_case2}--\eqref{constr_a<-3_gamma=-2} (except that the interval $-3<\a<-2$ appearing there should be replaced by $-3<\a<-\frac{5}{2}$ here). In this case $\bl$ can be an aligned null direction of the Maxwell field.

In the special subcase $F_{ij}^{(-2)}=0$ one instead has the same results as for case~\eqref{n>5_subrad} of section~\ref{subsubsec_a<-2_subcase} for $-3\le\a<\frac{5}{2}$, and those of case~\eqref{n>5_Coulomb} of section~\ref{subsubsec_a<-2_subcase} for $\a<-3$.
The latter includes asymptotically Coulombian fields.

\section{Full asymptotic behaviour for $R\neq0$} 

\label{sec_R}

As in section~\ref{R=0} we refer to appendix~\ref{app_setup} for technical details. Here we present the final results for the case $R\neq 0$, and further observe that now there is no need to treat $n=5$ separately.

\begin{enumerate}[(a)]

\item The ``generic'' behaviour is

\beqn
			& & F_{0i}=\frac{F_{0i}^{(-3)}}{r^3}+\ldots , \label{+1_R_generic} \\
			& & F_{01}=\frac{F_{01}^{(-3)}}{r^3}+\ldots  , \qquad F_{ij}=\frac{F_{ij}^{(-2)}}{r^2}+\ldots , \label{0_R_generic} \\
			& & F_{1i}=\frac{\tilde R}{2}\frac{F_{0i}^{(-3)}}{r}+\ldots , \label{-1_R_generic} 
\eeqn
where $F_{01}^{(-3)}$ is determined by \eqref{case2_a=-3} (except for $n=5$), and \eqref{B2} gives
\be
		\xi^{A0}_{[k}F_{ij],A}^{(-2)}+2F_{[ij|}^{(-2)}l_{1|k]}+2F_{l[j}^{(-2)}\m{l}{i}{k]}=-2\tilde R F_{0[j}^{(-3)}b_{ik]} .
\ee

The leading asymptotic term is of type N.

\item\label{R_a<-3} If $F_{0i}^{(-3)}=0$ then \eqref{+1_R_generic}--\eqref{-1_R_generic} reduce to
\beqn
			& & F_{0i}=\frac{F_{0i}^{(-4)}}{r^4}+\ldots , \label{+1_R_b} \\
			& & F_{01}=\frac{F_{01}^{(-3)}}{r^3}+\ldots , \qquad F_{ij}=\frac{F_{ij}^{(-2)}}{r^2}+\ldots ,  \\
			& & F_{1i}=\frac{F_{1i}^{(-2)}}{r^2}+\ldots , \label{-1_R_b} 
\eeqn
with \eqref{case2_a<-3} and (by combining \eqref{-C1+A2} and \eqref{+C1+A2})
\beqn
			& & F_{1i}^{(-2)}=-F_{ij}^{(-2)}l_{j1}-\frac{\tilde R}{2}F_{0i}^{(-4)} , \\
			& & -(n-5)\tilde R F_{0i}^{(-4)}=\xi^{A0}_jF_{ij,A}^{(-2)}+F_{ij}^{(-2)}(2l_{1j}+\m{j}{k}{k})+F_{jk}^{(-2)}\m{j}{i}{k}+(n-6)F_{ij}^{(-2)}l_{j1} .
\eeqn
Here \eqref{B2} gives \eqref{B2_case2}. 

The leading term is of type II. In this case $\a<4$ (i.e., $F_{0i}^{(-4)}=0$) is also permitted and, in particular, $\bl$ can be aligned with $F_{ab}$. As in the corresponding case with $\tilde R=0$ (eqs.~\eqref{+1_gamma=-2}--\eqref{-1_gamma=-2}), cf. (the weak-field limit of) certain solutions of \cite{OrtPodZof08} for examples.

\item If instead $F_{ij}^{(-2)}=0$, i.e., the magnetic term vanishes,  then \eqref{+1_R_generic}--\eqref{-1_R_generic} are replaced by
\beqn
			& & F_{0i}=\frac{F_{0i}^{(-3)}}{r^3}+\ldots , \label{+1_R_c} \\
			& & F_{01}=\frac{F_{01}^{(-3)}}{r^3}+\ldots , \qquad F_{ij}=\frac{F_{ij}^{(-3)}}{r^3}+\ldots , \\
			& & F_{1i}=\frac{\tilde R}{2}\frac{F_{0i}^{(-3)}}{r}+\ldots , \label{-1_R_c} 
\eeqn
with $F_{01}^{(-3)}$ and $F_{ij}^{(-3)}$ determined by \eqref{F01_gener}, \eqref{Fij_gener} with $\a=-3$ (except that $F_{01}^{(-3)}$ remains arbitrary for $n=5$). However, \eqref{Fij_gener} together with \eqref{D2} can be rearranged as 
\beqn
 & & F_{ij}^{(-3)}=2F_{0[i}^{(-3)}l_{j]1} , \\
 & & \xi^{A0}_{[j|}F_{0|i],A}^{(-3)}+2F_{0[i|}^{(-3)}l_{1|j]}+F_{0k}^{(-3)}\m{k}{[i}{j]}-F_{0[i}^{(-3)}l_{j]1}=0 , 
\eeqn
while \eqref{B2} gives
\be
		\tilde R F_{0[j}^{(-3)}b_{ik]}=0 .
\ee

Here the leading term is of type N.

\item If both $F_{0i}^{(-3)}=0$ and $F_{ij}^{(-2)}=0$ then we have
\beqn
			& & F_{0i}=\frac{F_{0i}^{(1-n)}}{r^{n-1}}+\ldots , \label{+1_R_coulomb} \\
			& & F_{01}=\frac{F_{01}^{(2-n)}}{r^{n-2}}+\ldots , \qquad F_{ij}=\frac{F_{ij}^{(1-n)}}{r^{n-1}}+\ldots  , \label{0_R_coulomb} \\
			& & F_{1i}=-\frac{\tilde R}{2}\frac{F_{0i}^{(1-n)}}{r^{n-3}}+\ldots , \label{-1_R_coulomb} 
\eeqn
where $F_{ij}^{(1-n)}$ is determined by \eqref{case2B_a=1-n}. 

The leading term is of type N.

\item\label{R_a<1-n} If (in addition to $F_{0i}^{(-3)}=0$ and $F_{ij}^{(-2)}=0$) also $F_{0i}^{(1-n)}=0$ then \eqref{+1_R_coulomb}--\eqref{-1_R_coulomb}  reduce to
\beqn
			& & F_{0i}=\frac{F_{0i}^{(-n)}}{r^{n}}+\ldots , \label{+1_R_coulomb_spec} \\
			& & F_{01}=\frac{F_{01}^{(2-n)}}{r^{n-2}}+\ldots , \qquad F_{ij}=\frac{F_{ij}^{(1-n)}}{r^{n-1}}+\ldots , \label{0_R_coulomb_spec} \\
			& & F_{1i}=\frac{F_{1i}^{(2-n)}}{r^{n-2}}+\ldots , \label{-1_R_coulomb_spec} 
\eeqn
with \eqref{case2B_a<1-n} and (by combining \eqref{-C1+A2} and \eqref{+C1+A2})
\beqn
			& & F_{1i}^{(2-n)}=F_{01}^{(2-n)}l_{i1}+\frac{\tilde R}{2}F_{0i}^{(-n)} , \label{F1i_R_coulomb_spec} \\
			& & (n-3)\tilde R F_{0i}^{(-n)}=\xi^{A0}_iF_{01,A}^{(2-n)}+(n-2)F_{01}^{(2-n)}(l_{1i}-l_{i1}) . \label{F0i_R_coulomb_spec}
\eeqn

Eq.~\eqref{B2} gives an equation containing the derivatives $\xi^{A0}_{[k}F_{ij],A}^{(1-n)}$, where, however, \eqref{case2B_a<1-n} (together with \eqref{F1i_R_coulomb_spec} and \eqref{F0i_R_coulomb_spec}) should be substituted -- this is not very useful for the purposes of the present paper and we shall therefore not present it. 

In this case the leading term is of type II. Values $\a<-n$ (i.e., $F_{0i}^{(-n)}=0$) are also permitted and, in particular, $\bl$ can be aligned with $F_{ab}$. Higher dimensional spinning AdS black holes carrying a ``small'' charge \cite{Aliev07} have the above behaviour.

\item If instead $F_{01}^{(2-n)}=0$, i.e., the Coulomb term vanishes (in addition to $F_{0i}^{(-3)}=0$ and $F_{ij}^{(-2)}=0$) then \eqref{+1_R_coulomb}--\eqref{-1_R_coulomb} are replaced by
\beqn
			& & F_{0i}=\frac{F_{0i}^{(1-n)}}{r^{n-1}}+\ldots , \label{+1_R_f} \\
			& & F_{01}=\frac{F_{01}^{(1-n)}}{r^{n-1}}+\ldots , \qquad F_{ij}=\frac{F_{ij}^{(1-n)}}{r^{n-1}}+\ldots , \\
			& & F_{1i}=-\frac{\tilde R}{2}\frac{F_{0i}^{(1-n)}}{r^{n-3}}+\ldots ,  \label{-1_R_f}
\eeqn
with $F_{01}^{(1-n)}$ and $F_{ij}^{(1-n)}$ determined by \eqref{F01_gener}, \eqref{Fij_gener} with $\a=1-n$. However, \eqref{F01_gener} together with \eqref{B1} can be rearranged as 
\beqn
 & & F_{01}^{(1-n)}=F_{0i}^{(1-n)}l_{i1} , \\
 & & \xi^{A0}_iF_{0i,A}^{(1-n)}+(n-2)F_{0i}^{(1-n)}l_{1i}+F_{0j}^{(1-n)}\m{j}{i}{i}-F_{0i}^{(1-n)}l_{i1}=0 .
\eeqn

The leading term is of type N. 

If, additionally, also $F_{0i}^{(1-n)}=0$ then one has $F_{ab}=0$ identically.

\end{enumerate}

Similarly as in the case $\tilde R=0$ (section~\ref{subsubsec_a<-2_subcase}) it follows that fields more special than type G must have (if $\bl$ is chosen to be aligned) $F_{ij}^{(-2)}\neq0$ (case~(\ref{R_a<-3})) or $F_{01}^{(2-n)}\neq0$ (case~(\ref{R_a<1-n})) --  in particular, type N fields having $\bl$ as an aligned null direction are not permitted. Note also that, as opposed to the case $\tilde R=0$, due to the presence of a cosmological constant the range of permitted values of $\a$ does not coincide with the set of real numbers and certain intervals are forbidden.

\section{Behaviour of $p$-form fields} 

\label{sec_p}

\subsection{General comments}

The full set of Maxwell equations was given in GHP notation in \cite{Durkeeetal10} for any $p$-form field $F_{a_1\ldots a_{p}}$ (we take $2\le p\le n-1$).\footnote{The integer $p$ of the present paper is $(p+1)$ in \cite{Durkeeetal10}.} The method used in the previous sections for the case $p=2$ can be similarly applied to the case of a generic $p$, after translating the equations of \cite{Durkeeetal10} into NP notation. As in \eqref{leading}, we assume that for $r\to\infty$
\beqn
 & & F_{0i_1\ldots i_{p-1}}=F_{0i_1\ldots i_{p-1}}^{(\a)}r^\a+\ldots , \nonumber \\ 
 & & F_{01i_1\ldots i_{p-2}}=F_{01i_1\ldots i_{p-2}}^{(\b)}r^\b+\ldots , \qquad F_{i_1\ldots i_{p}}=F_{i_1\ldots i_{p}}^{(\g)}r^\g+\ldots ,  \\ 
 & & F_{1i_1\ldots i_{p-1}}=F_{1i_1\ldots i_{p-1}}^{(\de)}r^\de+\ldots . \label{leading_evend} \nonumber
\eeqn
We do not want to investigate here again all possible cases (for various choices of boundary conditions)
but just make some general comments, especially concerning radiative fields. We will also point out a unique asymptotic behaviour for the special case $p=n/2$ in even spacetime dimension $n$, which is qualitatively different from that of the case $p=2$ when $n>4$ (and in fact of any other $p$). We assume the same Einstein spacetime background as in the previous sections, so that the expressions of section \ref{subsec_r_Ricci} for the Ricci rotation coefficients and derivative operators still apply.

We observe that in order to determine the asymptotic properties of the field $F_{a_1\ldots a_{p}}$ along $\bl$, the crucial terms to consider are those of order $\OOp{\b-1}$ in (3.3,\cite{Durkeeetal10}), i.e., $F_{01i_1\ldots i_{p-2}}^{(\b)}(\b+n-p)$, of order $\OOp{\g-1}$ in (3.4,\cite{Durkeeetal10}), i.e., $F_{i_1\ldots i_{p}}^{(\g)}(\g+p)$, and of order $\OOp{\de-1}$ in (3.5,\cite{Durkeeetal10})' and (3.5,\cite{Durkeeetal10}), i.e., $F_{1i_1\ldots i_{p-1}}^{(\de)}(2\de+n-2)$ and $F_{1i_1\ldots i_{p-1}}^{(\de)}(-2p+n)$.

In particular, it follows that when $\tilde R=0$ the boundary conditions $-\frac{n}{2}\le\a<1-\frac{n}{2}$ along with $F_{01i_1\ldots i_{p-2}}^{(p-n)}=0$ (for $p>\frac{n}{2}$) or $F_{i_1\ldots i_{p}}^{(-p)}=0$ (for $p<\frac{n}{2}$) lead again to the radiative case \eqref{+1_rad}--\eqref{-1_rad} (just add the appropriate number of spacelike indices in those formulas). In order for the radiative type~N term $F_{1i_1\ldots i_{p-1}}^{(1-\frac{n}{2})}$ in \eqref{-1_rad} to be non-zero, it is necessary that $F_{0i_1\ldots i_{p-1}}$ falls offs not faster than $1/r^{\frac{n}{2}}$ (except if $p=\frac{n}{2}$, see also \eqref{+1_n/2}--\eqref{-1_n/2} below), and for $\a=-\frac{n}{2}$ eq.~\eqref{+C1+A2_rad} generalizes to 
\be 
  (n-2p)F_{1i_1\ldots i_{p-1}}^{(1-\frac{n}{2})}=-2\left[X^{A0}F_{0i_1\ldots i_{p-1},A}^{(-\frac{n}{2})}+\left(\frac{n}{2}-1\right)F_{0i_1\ldots i_{p-1}}^{(-\frac{n}{2})}l_{11}+\sum_{s=1}^{p-1}F_{0i_1\ldots i_{s-1}ki_{s+1}\ldots i_{p-1}}^{(-\frac{n}{2})}\m{k}{i_s}{1}\right] .
\ee

A peeling-off as in \eqref{peeling_gen} also holds here, and the energy flux is given by $T_{11}\propto F_{1i_1\ldots i_{p-1}}F_{1i_1\ldots i_{p-1}}\sim 1/r^{n-2}$.

However, as already observed for the case $p=2$, the behaviour \eqref{+1_rad}--\eqref{-1_rad} is not permitted when $\tilde R\neq 0$. Instead, we note that if one looks for a possible (peeling-like) fall-off with $\delta>\b, \g$ (i.e., with b.w. $-1$ components $F_{1i_1\ldots i_{p-1}}$ falling off more slowly than b.w .0 components $F_{01i_1\ldots i_{p-2}}$ and $F_{i_1\ldots i_{p}}$), one finds from (3.5,\cite{Durkeeetal10}) and (3.5,\cite{Durkeeetal10})' that either $\a=-p-1$, $\de=-p+1$ or $\a=p-n-1$, $\de=p-n+1$. The further requirement $\delta=1-\frac{n}{2}>\b,\g$ (as in \eqref{-1_rad}) can be satisfied only when $p=\frac{n}{2}$, thus giving $\a=-1-\frac{n}{2}$, $\de=1-\frac{n}{2}$.  Eqs.~(3.3,\cite{Durkeeetal10}) and  (3.4,\cite{Durkeeetal10}) then further give $\b=-\frac{n}{2}=\g$. The same asymptotic behaviour follows also for $p=\frac{n}{2}$ with $\tilde R=0$ if one just assumes the boundary condition $\a=-1-\frac{n}{2}$ (this can be understood as a ``subcase'' of \eqref{+1_rad}--\eqref{-1_rad}).

\subsection{Asymptotic behaviour of $n/2$-forms (even dimensions)} 

\label{subsec_n/2}

In view of the above comments, for the special case $p=\frac{n}{2}$ in {\em even} dimensions we can thus present the unified asymptotic behaviour (for both $\tilde R=0$ and $\tilde R\neq0$)

\beqn
	& & F_{0i_1\ldots i_{p-1}}=\frac{F_{0i_1\ldots i_{p-1}}^{(-1-\frac{n}{2})}}{r^{\frac{n}{2}+1}}+\ldots , \label{+1_n/2} \\
	& & F_{01i_1\ldots i_{p-2}}=\frac{F_{01i_1\ldots i_{p-2}}^{(-\frac{n}{2})}}{r^{\frac{n}{2}}}+\ldots , \qquad F_{i_1\ldots i_{p}}=\frac{F_{i_1\ldots i_{p}}^{(-\frac{n}{2})}}{r^{\frac{n}{2}}}+\ldots , \qquad \left(p=\frac{n}{2}\right) \label{0_n/2} \\
	& & F_{1i_1\ldots i_{p-1}}=\frac{F_{1i_1\ldots i_{p-1}}^{(1-\frac{n}{2})}}{r^{\frac{n}{2}-1}}+\ldots . \label{-1_n/2}
\eeqn

The leading term is of type N and falls off as $1/r^{\frac{n}{2}-1}$, similarly as for $p=2$ with $\tilde R=0$, i.e. case~(\ref{n>5_rad}) of section~\ref{subsubsec_a<-2_subcase} (and as in the case of a generic $p$ mentioned above). However, as in the standard 4D case \eqref{Maxw4D} ($n=4$, $p=2$, $\a=-3$), and in contrast to the behaviour \eqref{+1_rad}--\eqref{-1_rad} for $p\neq\frac{n}{2}$, any of the terms in \eqref{+1_n/2}--\eqref{-1_n/2} can vanish without affecting the remaining expressions. In particular, algebraically special Maxwell fields of type N aligned with $\bl$ (i.e., with only \eqref{-1_n/2} being non-zero) are now permitted, cf. the 6D example of section~\ref{subsubsec_p=n/2_ex} below. Simple examples of type D aligned with $\bl$ and with $\bn$ (with only the terms \eqref{0_n/2} being non-zero and both falling off at the same speed) can also be constructed (section~\ref{subsubsec_p=n/2_ex}).

Furthermore, if we assume that also the subleading term of $F_{1i_1\ldots i_{p-1}}$ is power-like, it is easy to see from (3.5,\cite{Durkeeetal10})' that \eqref{-1_n/2} can be refined as $F_{1i_1\ldots i_{p-1}}=F_{1i_1\ldots i_{p-1}}^{(1-\frac{n}{2})}r^{1-\frac{n}{2}}+\OO{\frac{n}{2}}$. This implies the peeling-off
\be
	F_{a_1\ldots a_{p}}=\frac{N_{a_1\ldots a_{p}}}{r^{\frac{n}{2}-1}}+\frac{II_{a_1\ldots a_{p}}}{r^{\frac{n}{2}}}+\ldots \qquad\qquad \left(p=\frac{n}{2}\right) ,
	\label{peeling_n/2}
\ee
which is clearly qualitatively different from that of the case $p=2$ (cf.~\eqref{peeling_gen}), and also holds in the presence of a cosmological constant. For $n=4$ this agrees with the standard result \eqref{Maxw4D}.

Exact even-dimensional Robinson-Trautuman solutions coupled to $\frac{n}{2}$-forms have recently been obtained \cite{OrtPodZof15} that display the behavior \eqref{peeling_n/2} in the full Einstein-Maxwell theory. These include both type II and N Maxwell fields.

\subsubsection{Examples for $n=6$, $p=3$}

\label{subsubsec_p=n/2_ex}

To conclude, let us give two simple examples of 3-forms that solve Maxwell's equations in the background of a six-dimensional Schwarzschild-Tangherlini black hole (which is of Weyl type D). The corresponding metric can be written as

\be
 \d s^2=r^2P^{-2}\delta_{\hat i\hat j}\d x^{\hat i}\d x^{\hat j}+2\d u\d r-2H\d u^2 ,
 \label{RT}
\ee
with
\beqn 
  & & P=1+\frac{K}{4}\rho^2 , \qquad \rho^2=\sum_{\hat i=2}^5(x^{\hat i})^2 , \nonumber \\
	& & 2H=K-\frac{\Lambda}{10}\,r^2-\frac{\mu}{r^{3}} 	\qquad (K=0,\pm 1) ,
\eeqn 
where the hat over the indices $\hat i, \hat j, \ldots$ indicates that they are coordinate (and not frame) indices.

The ``null'' frame
\be
 \bl=\pa_r , \qquad \bn=\pa_u+H\pa_r , \qquad \bbm_{(i)}=\frac{P}{r}\pa_{\hat i} , \label{frame_6D}
\ee
is parallelly transported along the null geodesic vector field $\bl$ (which is expanding, twistfree and shearfree \cite{PodOrt06}), and will be used below to evaluate the frame field components. Both $\bl$ and $\bn$ are multiple WANDs \cite{PodOrt06}.

\paragraph{Type D field}

A solution of Maxwell's equations having only b.w. 0 components and both electric and magnetic terms is given by
\be
 F=Q_E\frac{P^2x^{\hat i}}{r^2\rho^4}\d u\wedge\d r\wedge\d x^{\hat i}+Q_M\frac{\epsilon_{\hat i\hat j\hat k\hat l}x^{\hat l}}{\rho^4}\d x^{\hat i}\wedge\d x^{\hat j}\wedge\d x^{\hat k} ,
\label{6D_typeD}
\ee
where $Q_E$ and $Q_M$ are two arbitrary constants parametrizing, respectively, the electric and magnetic field strength, $\epsilon_{\hat i\hat j\hat k\hat l}$ is the Levi-Civita symbol of the transverse 4-space of constant $u$ and $r$ (such that $\epsilon_{\hat 2\hat 3\hat 4\hat 5}=+1$), and summation over repeated indices is understood.

In the frame \eqref{frame_6D} the fall-off behaviour of \eqref{6D_typeD} is given by $F_{01i}\sim 1/r^3$, $F_{ijk}\sim 1/r^3$ (all other components being zero), in agreement with the general result \eqref{0_n/2}.

\paragraph{Type N field}

An example of a type N Maxwell field (following from a discussion of optical structures in Robsinson-Trautman spacetimes \cite{OrtPraPra13}) is given instead by 
\be
 F=K_{\hat i\hat j}(u)\d u\wedge\d x^{\hat i}\wedge\d x^{\hat j} ,
\label{6D_typeN}
\ee
where the $K_{\hat i\hat j}(u)$ can be arbitrary functions of $u$ (this is a partial extension of a type N 4D solution in a Minkowski background given in \cite{RobTra62}, and a subcase of a solution given in \cite{Trautman02b}).\footnote{The solution \eqref{6D_typeN} can in fact be straightforwardly generalized to any even $n$ dimension in the background \eqref{RT} (for arbitrary $n$ the last term of $H$ reads $-\mu/r^{n-3}$) by just adding the necessary number of 1-forms $\d x^{\hat i}$.} In the frame \eqref{frame_6D}, from \eqref{6D_typeN} one finds the only non-zero components $F_{1ij}\sim 1/r^2$, in agreement with \eqref{-1_n/2}. It may be interesting also to observe that the field \eqref{6D_typeN} is self-dual (or anti-self-dual) if its only non-zero components are $K_{23}=-K_{45}$ (or $K_{23}=+K_{45}$).

Furthermore, thanks to the linearity of Maxwell's equations one can obviously superimpose the solutions \eqref{6D_typeD} and \eqref{6D_typeN}, obtaining the peeling-off $F_{abc}=N_{abc}/r^2+D_{abc}/r^3$, which is a special instance of the general result \eqref{peeling_n/2} (in the present case all higher order terms vanish identically).

\section{Conclusions}

We have determined the asymptotic behaviour of test Maxwell fields near null infinity in a class of higher dimensional Einstein spacetimes (including in particular asymptotically flat spacetimes, and constant curvature spacetimes). The obtained possible decays are given for $\tilde R=0$ by \eqref{F0i_case1}--\eqref{F1i_case1}, \eqref{+1_gamma=-2}--\eqref{-1_gamma=-2}, \eqref{+1_rad}--\eqref{-1_rad} and \eqref{+1_n>5_Coulomb}--\eqref{-1_n>5_Coulomb} (but recall some additional special cases for $n=5$ in section~\ref{subsec_n=5}), and for $\tilde R\neq0$ by \eqref{+1_R_generic}--\eqref{-1_R_generic}, \eqref{+1_R_b}--\eqref{-1_R_b}, \eqref{+1_R_c}--\eqref{-1_R_c}, \eqref{+1_R_coulomb}--\eqref{-1_R_coulomb}, \eqref{+1_R_coulomb_spec}--\eqref{-1_R_coulomb_spec} and \eqref{+1_R_f}--\eqref{-1_R_f}. The full fall-off in general follows only once boundary conditions on field components of b.w. $+1$ {\em and} 0 are fixed. This is different from the situation in four dimensions and, correspondingly, results in a new peeling-off in higher dimensions. We did not assume a series expansion for the Maxwell field and stopped at the (sub)leading order. For that reason, we expect that several of the results of the present paper will apply also to the full Einstein-Maxwell theory, at the leading order. However, a careful analysis of the Bianchi/Ricci equations as done in \cite{OrtPra14}, but also taking into account the back-reaction of the Maxwell field, would be required to say when precisely that happens  (for example, ``uniform'' magnetic fields in the full theory are described by Melvin-like solutions \cite{Gibbons86,GibWil87,Ortaggio05}, which have an asymptotics different from their test-field limit mentioned in this paper). Note also that we have studied only the $r$-dependence, and not the full integrability, of the Maxwell equations. The latter would be necessary in order to rigorously prove that all the possible cases that arose indeed really exist, but it goes beyond the scope of this paper. However, in several cases we mentioned explicit examples falling into those classes of solutions, thus proving they are not empty.

We also discussed the case of general $p$-forms that solve the Maxwell equations. Perhaps not entirely surprisingly, the case $p=\frac{n}{2}$ in even dimensions turned out to be special, exhibiting a peeling-off similar to the standard four-dimensional case (both with and without a cosmological constant) and qualitatively different from that of any other $p$. Simple examples demonstrating how $p=\frac{n}{2}$ is special have been provided (see also \cite{OrtPodZof15}).

\section*{Acknowledgments}

I am grateful to Alena Pravdov\'a for reading the manuscript. Support from research plan {RVO: 67985840} and research grant GA\v{C}R 14-37086G is also acknowledged.

\renewcommand{\thesection}{\Alph{section}}
\setcounter{section}{0}

\renewcommand{\theequation}{{\thesection}\arabic{equation}}

\section{Definitions of the Ricci rotation coefficients}
\setcounter{equation}{0}

\label{app_Ricci}

In addition to \eqref{rho}, the remaining Ricci rotation coefficients are defined by (see \cite{OrtPraPra13rev} and references therein)
\beqn 
 & & \kappa_i\equiv L_{i0}=\ell_{a;b}m_{(i)}^a l^b  , \qquad L_{10}=\ell_{a;b}n^a l^b  , \qquad  L_{1i}=\ell_{a;b}n^a m_{(i)}^b  , \qquad L_{i1}=\ell_{a;b}m_{(i)}^a n^b 	, \qquad  L_{11}=\ell_{a;b}n^a n^b , 	\nonumber \\
 & & \M{i}{j}{0}=m_{(i)a;b}m_{(j)}^a l^b , \qquad \M{i}{j}{k}=m_{(i)a;b}m_{(j)}^a m_{(k)}^b , \qquad \M{i}{j}{1}=m_{(i)a;b}m_{(j)}^a n^b , \\
 & & N_{i0}=n_{a;b}m_{(i)}^a l^b  , \qquad N_{ij}=n_{a;b}m_{(i)}^a m_{(j)}^b  , \qquad  N_{i1}=n_{a;b}m_{(i)}^a n^b  . \nonumber 
\eeqn

\section{Technical details of the derivation of the $r$-dependence (case $p=2$)}
\setcounter{equation}{0}

\label{app_setup}



\subsection{Maxwell equations in a parallelly transported frame} 

In the frame defined in section~\ref{subsec_notat}, with the defitions \eqref{der_oper} and with the conditions \eqref{l_geod} and \eqref{parall_frame}, the empty-space Maxwell equations $F^a_{\ \; b;a}=0$ and $F_{[ab;c]}=0$ take the form (ordered by b.w.)

\beqn
	& & DF_{01}+F_{01}\R{}+\delta_iF_{0i}-F_{0i}L_{1i}+F_{0j}\M{j}{i}{i}-F_{ij}\R{ij}=0 , \label{A1} \\
	& & DF_{ij}+{2F_{[i|k}\R{k|j]}}+2\delta_{[j|}F_{0|i]}-{2F_{0[i|}L_{1|j]}}+2F_{0k}\M{k}{[i}{j]}+2F_{01}\R{[ij]}=0 , \label{C2} \\
	& & 2DF_{1i}+F_{1i}\R{}-2F_{1j}\R{[ij]}-\delta_jF_{ij}-F_{jk}\M{j}{i}{k}+F_{ji}\M{j}{k}{k}+\delta_iF_{01}-F_{0j}(N_{ij}+N_{ji})+F_{0i}N_{jj}=0 , \label{-C1+A2} \\
	& & -2\Delta F_{0i}+2F_{0i}L_{11}-F_{0i}N_{jj}-2F_{0j}(\M{j}{i}{1}-N_{[ij]})+\delta_j F_{ij}-2F_{ij}L_{j1}+F_{jk}\M{j}{i}{k}-F_{ji}\M{j}{k}{k} \nonumber \\
	& & \qquad\qquad\qquad\qquad\qquad {}+\delta_iF_{01}-2F_{01}L_{i1}+F_{1j}(\R{ij}+\R{ji})-F_{1i}\R{}=0 , \label{+C1+A2} \\
	& & \delta_{[k}F_{ij]}+2F_{l[j}\M{l}{i}{k]}+2F_{0[j}N_{ik]}+2F_{1[j}\R{ik]}=0 , \label{B2} \\
	& & -\Delta F_{01}-F_{01}N_{ii}+\delta_iF_{1i}+F_{1i}(L_{1i}-L_{i1})+F_{1j}\M{j}{i}{i}+F_{0i}N_{i1}-F_{ij}N_{ij}=0 , \label{B1} \\
	& & \Delta F_{ij}-2F_{k[i|}(N_{k|j]}+\M{k}{|j]}{1})+2\delta_{[j|}F_{1|i]}+2F_{1[i|}(L_{1|j]}-L_{|j]1})+2F_{1k}\M{k}{[i}{j]} \nonumber \\
	& & \qquad\qquad\qquad\qquad\qquad {}-2F_{0[i}N_{j]1}-2F_{01}N_{[ij]}=0 . \label{D2} 
\eeqn

These can also be obtained by translating the corresponding GHP equations \cite{Durkeeetal10} into the NP notation. 
Note that \eqref{-C1+A2} and \eqref{+C1+A2} are linear combinations of two b.w. 0 equations coming, respectively, from $F^a_{\ \; b;a}=0$ and $F_{[ab;c]}=0$. This choice will be convenient for practical purposes since separates the derivative terms $DF_{1i}$ and $\Delta F_{0i}$ into two different equations. Eq.~\eqref{B2} becomes a trivial identity in 4D because of the total antisymmetrization.

\subsection{Asymptotic behaviour of b.w. 0 components}

\label{subsec_bw0}

Let us start with \eqref{A1} and \eqref{C2}. After substituting the first three of \eqref{leading}, by examining the leading order terms one concludes that the possible asymptotic behaviors generically are the following.
\begin{enumerate}
	\item If $\a\ge-2$ then 
	\be
		\g=\b=\a ,
	\ee	
	and
	\beqn
	 & & (\a+n-2)F_{01}^{(\a)}=-\xi^{A0}_iF_{0i,A}^{(\a)}+(\a+1)F_{0i}^{(\a)}l_{1i}-F_{0j}^{(\a)}\m{j}{i}{i} , \label{F01_gener} \\
	 & & (\a+2)F_{ij}^{(\a)}=-2\xi^{A0}_{[j|}F_{0|i],A}^{(\a)}+2(\a+1)F_{0[i|}^{(\a)}l_{1|j]}-2F_{0k}^{(\a)}\m{k}{[i}{j]} . \label{Fij_gener} 
	\eeqn
	Clearly $F_{ij}^{(\a)}$ is undetermined in the case $\a=-2$.\footnote{For $\a=-2$ eq.~\eqref{Fij_gener} becomes a constraint on $F_{0i}^{(\a)}$. Comparison with the 4D analysis of \cite{Kroon00} suggests that in this special case a more general framework to consider may be that of polyhomogenous expansions. This however goes beyond the scope of this paper, where we restrict to power-like leading terms (cf.~\eqref{leading}). (A similar comment applies to case (A) below when $\a=2-n$.)}

	\item If $\a<-2$ and $n>4$ then 
	  \be 
			\g=-2>\b , 
		\ee	
		with $F_{ij}^{(-2)}$ being integration functions. Thanks to \eqref{L_2} we have $F_{ij}\R{ij}=\OO{4}$, and \eqref{A1} thus gives $\b=\max\{-3,\a\}$, so that
		\begin{itemize}
		\item if $-3<\a<-2$: eq.~\eqref{F01_gener} holds 
					
		\item if $\a=-3$:
				\be
					(n-5)F_{01}^{(-3)}=-\xi^{A0}_iF_{0i,A}^{(-3)}-2F_{0i}^{(-3)}l_{1i}-F_{0j}^{(-3)}\m{j}{i}{i}+F_{ij}^{(-2)}b_{ij} 
					\label{case2_a=-3}
				\ee
		\item if $\a<-3$:
				\be
					(n-5)F_{01}^{(-3)}=F_{ij}^{(-2)}b_{ij} .
					\label{case2_a<-3}
				\ee
	
	\end{itemize}

		In the special case $F_{ij}^{(-2)}=0$, further possible subcases are:
		\begin{enumerate}[(A)]
			\item if $2-n\le\a<-2$ then $\g=\b=\a$ and \eqref{F01_gener}, \eqref{Fij_gener} still apply;
			\item if $\a<2-n$ then $\b=2-n>\g$, with the integration function $F_{01}^{(2-n)}$. Using $F_{01}\R{[ij]}=\OO{n}$, \eqref{C2} gives $\g=\max\{1-n,\a\}$, so that
		\begin{itemize}
		\item if $1-n<\a<2-n$: eq.~\eqref{Fij_gener} holds 
					
		\item if $\a=1-n$:
				\be
					(3-n)F_{ij}^{(1-n)}=-2\xi^{A0}_{[j|}F_{0|i],A}^{(1-n)}+2(2-n)F_{0[i|}^{(1-n)}l_{1|j]}-2F_{0k}^{(1-n)}\m{k}{[i}{j]}-2F_{01}^{(2-n)}b_{[ij]} 
					\label{case2B_a=1-n}
				\ee
		\item if $\a<1-n$:
				\be
					(3-n)F_{ij}^{(1-n)}=-2F_{01}^{(2-n)}b_{[ij]} .
					\label{case2B_a<1-n}
				\ee
	
	\end{itemize}
		\end{enumerate}	
		These results apply also for $n=4$, but only as a special subcase of a more general behaviour (see the next point below).
  \item If $\a<-2$ and $n=4$, generically one has
	 \be
		\b=\g=-2 , 
	 \ee	
	with two integration functions $F_{01}^{(-2)}$ and $F_{ij}^{(-2)}$. Special subcases with $F_{01}^{(-2)}=0$ and/or $F_{ij}^{(-2)}=0$ are possible. The fall-off of the remaining components in four dimensions is given in appendix~\ref{app_4D}.		

\end{enumerate}

The next step is to determine the $r$-dependence of the b.w. $-1$ components $F_{1i}$ (and to derive certain constraints on integration functions). However, in view of sections \ref{Ricci_subsubsec_R=0} and \ref{Ricci_subsubsec_R}, in all the remaining Maxwell equations some of the leading terms differ in the two cases $\tilde R=0$ and $\tilde R\neq0$, which thus need to be treated separately.

\subsection{Full asymptotic behaviour for $R=0$: sketch of the procedure}

\label{R=0_sketch}

In general, the leading order term of \eqref{-C1+A2} can be of order $\OOp{\a-1}$, $\OOp{\b-1}$, $\OOp{\g-1}$ or $\OOp{\de-1}$, depending on the relative value of the parameters $\a$, $\b$, $\g$, $\de$. In particular, at the leading order $F_{1i}$ gives a term $(2\de+n-2)F_{1i}^{(\de)}r^{\de-1}$. It is thus clear that either $\de\le\max\{\a,\b,\g\}$ or $\de=1-\frac{n}{2}$. On the other hand, the leading order term of \eqref{+C1+A2} can be of order $\OOp{\a}$, $\OOp{\b-1}$, $\OOp{\g-1}$ or $\OOp{\de-1}$, the latter originating from $(4-n)F_{1i}^{(\de)}r^{\de-1}$. By combining the conditions from \eqref{-C1+A2} and \eqref{+C1+A2} we obtain that either: 
\begin{enumerate}[(i)]
	\item $\de=\max\{\a,\b,\g\}$ ($\de$ can be also smaller than this, which we consider implicitly as a ``subcase'') 
	\item $\de=1-\frac{n}{2}$ with $-\frac{n}{2}\le\a<1-\frac{n}{2}$ and $\b,\g<1-\frac{n}{2}$. 
\end {enumerate}

In turn, this result must be intersected with the conditions obtained in section~\ref{subsec_bw0}. Case (i) is compatible with both cases 1. (giving $\de=\a$) and 2. (giving $\de=-2$) of section~\ref{subsec_bw0} -- in this case \eqref{-C1+A2} determines $F_{1i}^{(\de)}$ (except when $\max\{\a,\b,\g\}=1-\frac{n}{2}$, see section~\ref{R=0} and appendix~\ref{app_transv_R=0} for details) and then \eqref{+C1+A2} gives a constraint.
Case (ii) distinguishes between $n=5$ and $n\ge6$: for $n=5$ it falls into case 1. for $-2\le\a<-\frac{3}{2}$ and in case 2. (possibly, 2.(A) if $F_{ij}^{(-2)}=0$) for $-\frac{5}{2}\le\a<-2$, while for $n\ge6$ it must belong to case 2.(A). In case (ii) eq.~\eqref{-C1+A2} is identically satisfied at the leading order $\OO{\frac{n}{2}}$, while the leading order $\OOp{\a}$ of \eqref{+C1+A2} either gives a constraint on $F_{0i}^{(\a)}$ (for $-\frac{n}{2}<\a<1-\frac{n}{2}$) or determines $F_{1i}^{(1-\frac{n}{2})}$ in terms of $F_{0i}^{(-\frac{n}{2})}$ (for $\a=-\frac{n}{2}$). 
The remaining Maxwell equations \eqref{B2}--\eqref{D2} give constraints. 

It is clear that, due to the term $(4-n)F_{1i}^{(\de)}r^{\de-1}$ in \eqref{+C1+A2}, the above discussion does not apply to the 4D case: when $n=4$, if $\de=-1$ then $\a,\b,\g$ can be arbitrarily small, cf. appendix~\ref{app_4D}.
All the thus obtained results are summarized in section~\ref{R=0} in the various possible cases, treated separately for $n\ge6$ and  $n=5$.

\subsection{Full asymptotic behaviour for $R\neq 0$: sketch of the procedure}

The method used for $\tilde R\neq0$ is essentially the same as the one of section~\ref{R=0_sketch}. It will thus suffice here to emphasize the main differences. We note in particular that, due to \eqref{Ricci_R} and \eqref{U_R}, now both \eqref{-C1+A2} and \eqref{+C1+A2} contain terms of order $\OOp{\a+1}$ (in addition to $\OOp{\b-1}$, $\OOp{\g-1}$, $\OOp{\de-1}$, which keep the same form as when $\tilde R=0$), as opposed to the case $\tilde R=0$ (cf. the corresponding comments in section~\ref{R=0_sketch}). These are given, respectively, by
$\frac{\tilde R}{2}(4-n)F_{0i}^{(\a)}$ and $\frac{\tilde R}{2}(n+2+2\a)F_{0i}^{(\a)}$. It is thus immediately clear that $\a\le\max\{\b-2,\g-2,\de-2\}$ (unless $n=4$ and $\a=-3$, see appendix~\ref{app_4D}), and that we cannot have $\de>\b,\g$ unless $\de=\a+2$. In the latter case, however, \eqref{-C1+A2} and \eqref{+C1+A2} reveal that the only possibilities are either $\a=-3$, $\de=-1$ with $F_{1i}^{(-1)}=\frac{\tilde R}{2}F_{0i}^{(-3)}$, or $\a=1-n$, $\de=3-n$ with $F_{1i}^{(3-n)}=-\frac{\tilde R}{2}F_{0i}^{(1-n)}$. These observations imply that case 1. ($\a\ge-2$) of section~\ref{subsec_bw0} is thus forbidden when $\tilde R\neq0$, and in fact necessarily $\a\le-3$ here. Further using the results of section~\ref{subsec_bw0} and the Maxwell equations \eqref{-C1+A2}--\eqref{D2} one readily arrives at the possible behaviours summarized in section~\ref{sec_R} (we observe that in several cases some of the eqs.~\eqref{B2}--\eqref{D2} are identically satisfied at the leading order, and thus will not appear in section~\ref{sec_R}, nor a separate appendix will be now necessary).

\section{Non-radial Maxwell equations in the case $R=0$ ($p=2$)}
\setcounter{equation}{0}

\label{app_transv_R=0}

We give here the ``transverse'' Maxwell equations for the cases described in section~\ref{subsec_nge6}. Although there such equations are already referred to where appropriate, for further clarity we keep here for subsections the same titles as those used in section~\ref{subsec_nge6}.

\subsection{Fall-off of $F_{0i}$ with $\a\ge-2$}

	For \eqref{F0i_case1}--\eqref{F1i_case1}, from \eqref{+C1+A2}--\eqref{D2} one obtains   
	\beqn
		& & X^{A0}F_{0i,A}^{(\a)}-{F_{0i}^{(\a)}}l_{11}(\a+1)+F_{0j}^{(\a)}\m{j}{i}{1}=0 , \label{+C1+A2_gener} \\
		& & \xi^{A0}_{[k}F_{ij],A}^{(\a)}-\a F_{[ij|}^{(\a)}l_{1|k]}+2F_{l[j}^{(\a)}\m{l}{i}{k]}+2F_{0[j}^{(\a)}n_{ik]}=0 , \label{B2_case1} \\
		& & X^{A0}F_{01,A}^{(\a)}-\a F_{01}^{(\a)}l_{11}-F_{0i}^{(\a)}n_{i1}=0 , \label{B1_case1} \\
		& & X^{A0}F_{ij,A}^{(\a)}-\a F_{ij}^{(\a)}l_{11}-2F_{k[i}^{(\a)}\m{k}{j]}{1}-2F_{0[i}^{(\a)}n_{j]1}=0 . \label{D2_case1}
	\eeqn
	
	In all the above equations the expressions \eqref{F01_gener} and \eqref{Fij_gener} can of course be substituted, if desired. (The same results apply in certain subcases with $\a<-2$ mentioned below.)

\subsection{Fall-off of $F_{0i}$ with $\a<-2$: generic case}

For \eqref{+1_gamma=-2}--\eqref{-1_gamma=-2}, for all values $\a<-2$ we have (from \eqref{B2} and \eqref{D2}) 
 \beqn
		& & \xi^{A0}_{[k}F_{ij],A}^{(-2)}+2F_{[ij|}^{(-2)}l_{1|k]}+2F_{l[j}^{(-2)}\m{l}{i}{k]}=0 , \label{B2_case2} \\
		& & X^{A0}F_{ij,A}^{(-2)}+2F_{ij}^{(-2)}l_{11}-2F_{k[i}^{(-2)}\m{k}{j]}{1}=0 \label{D2_case2} .
\eeqn

	Further conditions depend on the value of $\a$ because of the $\OOp{\a}$ terms in \eqref{+C1+A2} and \eqref{B1}.

	\begin{itemize}
		\item If $-3<\a<-2$, \eqref{-C1+A2} gives
				\be
					(n-6)F_{1i}^{(-2)}=\xi^{A0}_jF_{ij,A}^{(-2)}+F_{ij}^{(-2)}(2l_{1j}+\m{j}{k}{k})+F_{jk}^{(-2)}\m{j}{i}{k} , \label{-C1+A2_gamma=-2} 
				\ee		
			and \eqref{+C1+A2}, \eqref{B1} give, respectively, \eqref{+C1+A2_gener} and \eqref{B1_case1}. $F_{1i}^{(-2)}$ is thus here undetermined in the special case $n=6$.

		\item If $\a=-3$, \eqref{-C1+A2} and \eqref{+C1+A2} can be combined to obtain 
				\beqn
					& & F_{1i}^{(-2)}=-F_{ij}^{(-2)}l_{j1}-\left(X^{A0}F_{0i,A}^{(-3)}+2F_{0i}^{(-3)}l_{11}+F_{0j}^{(-3)}\m{j}{i}{1}\right) , \label{F1i_a=-3_gamma=-2} \\
				  & & \xi^{A0}_jF_{ij,A}^{(-2)}+F_{ij}^{(-2)}(2l_{1j}+\m{j}{k}{k})+F_{jk}^{(-2)}\m{j}{i}{k} \nonumber \\
					& & \qquad\qquad\qquad {}+(n-6)\left[F_{ij}^{(-2)}l_{j1}+\left(X^{A0}F_{0i,A}^{(-3)}+2l_{11}F_{0i}^{(-3)}+F_{0j}^{(-3)}\m{j}{i}{1}\right)\right]=0 , \label{constr_a=-3_gamma=-2}
				\eeqn
				while \eqref{B1} gives an equation containing the derivatives $X^{A0}F_{01,A}^{(-3)}$ and $\xi^{A0}_jF_{1i,A}^{(-2)}$ (where, however, \eqref{case2_a=-3} and \eqref{F1i_a=-3_gamma=-2} should be substituted), which we omit.
			
		\item If $\a<-3$, we get the same equations as for the previous case $\a=-3$, except that the now subleading terms $F_{0i}^{(\a)}$ disappear, i.e.,
				\beqn
					& & F_{1i}^{(-2)}=-F_{ij}^{(-2)}l_{j1} , \\
					& & \xi^{A0}_jF_{ij,A}^{(-2)}+F_{ij}^{(-2)}(2l_{1j}+\m{j}{k}{k})+F_{jk}^{(-2)}\m{j}{i}{k}+(n-6)F_{ij}^{(-2)}l_{j1}=0 . \label{constr_a<-3_gamma=-2}
				\eeqn

	\end{itemize}

\subsection{Fall-off of $F_{0i}$ with $\a<-2$: subcase $F_{ij}^{(-2)}=0$ (case $-\frac{n}{2}\le\a<1-\frac{n}{2}$)}

\begin{enumerate}[(a)]

			\item See point (\ref{n>5_gen}) of section~\ref{subsubsec_a<-2_subcase}.
			
			\item Here only equations for the case \eqref{+1_rad}--\eqref{-1_rad} with $\a=-\frac{n}{2}$ need to be given (cf. point (\ref{n>5_rad}) in section~\ref{subsubsec_a<-2_subcase}). These are (from \eqref{B1}, \eqref{D2}, \eqref{B2})
				\beqn	
					& & X^{A0}F_{01,A}^{(-\frac{n}{2})}+\frac{n}{2}F_{01}^{(-\frac{n}{2})}l_{11}-F_{0i}^{(-\frac{n}{2})}n_{i1} \nonumber \\
					& & \qquad\qquad {}-\xi^{A0}_{i}F_{1i,A}^{(1-\frac{n}{2})}-F_{1i}^{(1-\frac{n}{2})}\left(\frac{n}{2}l_{1i}-l_{i1}\right)-F_{1j}^{(1-\frac{n}{2})}\m{j}{i}{i}=0 , \label{B1_case2_rad} \\
					& & X^{A0}F_{ij,A}^{(-\frac{n}{2})}+\frac{n}{2}F_{ij}^{(-\frac{n}{2})}l_{11}-2F_{k[i}^{(-\frac{n}{2})}\m{k}{j]}{1}-2F_{0[i}^{(-\frac{n}{2})}n_{j]1} \nonumber \qquad\qquad\qquad\qquad \left(\a=-\frac{n}{2}\right) \\
					& & \qquad\qquad {}+2\xi^{A0}_{[j|}F_{1|i],A}^{(1-\frac{n}{2})}+2F_{1[i|}^{(1-\frac{n}{2})}\left(\frac{n}{2}l_{1|j]}-l_{|j]1}\right)+2F_{1k}^{(1-\frac{n}{2})}\m{k}{[i}{j]}=0 , \label{D2_case2_rad} \\
					& & \xi^{A0}_{[k}F_{ij],A}^{(-\frac{n}{2})}+\frac{n}{2}F_{[ij|}^{(-\frac{n}{2})}l_{1|k]}+2F_{l[j}^{(-\frac{n}{2})}\m{l}{i}{k]}+2F_{0[j}^{(-\frac{n}{2})}n_{ik]}+2F_{1[j}^{(1-\frac{n}{2})}b_{ik]}=0  . \label{B2_case2_rad}
				\eeqn

			\item See point (\ref{n>5_subrad}) of section~\ref{subsubsec_a<-2_subcase}.
			
			\item For all values $\a<2-n$ we have from \eqref{B1} 
	\be
	 X^{A0}F_{01,A}^{(2-n)}+(n-2)F_{01}^{(2-n)}l_{11}=0 . \label{case2B_a<1-n_first}
	\ee
			
   The conditions following from \eqref{+C1+A2}, \eqref{B2} and \eqref{D2} depend on $\a$ (\eqref{-C1+A2_case2_a<2-n} is also used in \eqref{mixed_case2_a<2-n} and \eqref{mixed2_case2_a<2-n}).

			\begin{itemize}
					\item If $1-n<\a<2-n$: eqs.~\eqref{+C1+A2_gener}, \eqref{B2_case1} and \eqref{D2_case1} hold.
					
					\item If $\a=1-n$: 
						\beqn
						  & & \xi^{A0}_iF_{01,A}^{(2-n)}+(n-2)F_{01}^{(2-n)}(l_{1i}-l_{i1}) \nonumber \\
							& & \qquad\qquad {}-(n-2)\left[X^{A0}F_{0i,A}^{(1-n)}+(n-2)F_{0i}^{(1-n)}l_{11}+F_{0j}^{(1-n)}\m{j}{i}{1}\right]=0 , \label{mixed_case2_a<2-n} \\
			        & & \xi^{A0}_{[k}F_{ij],A}^{(1-n)}+(n-1)F_{[ij|}^{(1-n)}l_{1|k]}+2F_{l[j}^{(1-n)}\m{l}{i}{k]}+2F_{1[j}^{(2-n)}b_{ik]}+2F_{0[j}^{(1-n)}n_{ik]}=0 , \\ 
							& &  X^{A0}F_{ij,A}^{(1-n)}+(n-1)F_{ij}^{(1-n)}l_{11}-2F_{k[i}^{(1-n)}\m{k}{j]}{1}-2F_{01}^{(2-n)}n_{[ij]}-2F_{0[i}^{(1-n)}n_{j]1} \nonumber  \\
						  & &  \qquad\qquad {}+2\xi^{A0}_{[j|}F_{1|i],A}^{(2-n)}+2F_{1[i|}^{(2-n)}\left[(n-1)l_{1|j]}-l_{|j]1}\right]+2F_{1k}^{(2-n)}\m{k}{[i}{j]}=0 ,   
						\eeqn
						with \eqref{case2B_a=1-n}.
					
					\item If $\a<1-n$: one has the same equations as for the previous case $\a=1-n$, but without the now subleading terms $F_{0i}^{(\a)}$, i.e.,	
						\beqn
						  & & \xi^{A0}_iF_{01,A}^{(2-n)}+(n-2)F_{01}^{(2-n)}(l_{1i}-l_{i1})=0 , \label{mixed2_case2_a<2-n} \\
							& & \xi^{A0}_{[k}F_{ij],A}^{(1-n)}+(n-1)F_{[ij|}^{(1-n)}l_{1|k]}+2F_{l[j}^{(1-n)}\m{l}{i}{k]}+2F_{1[j}^{(2-n)}b_{ik]}=0 , \\ 
							& &  X^{A0}F_{ij,A}^{(1-n)}+(n-1)F_{ij}^{(1-n)}l_{11}-2F_{k[i}^{(1-n)}\m{k}{j]}{1}-2F_{01}^{(2-n)}n_{[ij]} \nonumber \\
						  & &  \qquad\qquad {}+2\xi^{A0}_{[j|}F_{1|i],A}^{(2-n)}+2F_{1[i|}^{(2-n)}\left[(n-1)l_{1|j]}-l_{|j]1}\right]+2F_{1k}^{(2-n)}\m{k}{[i}{j]}=0 ,  \label{case2B_a<1-n_last}
						\eeqn					
						with \eqref{case2B_a<1-n}.
			\end{itemize}

\end{enumerate}

\section{Asymptotic behaviour for $n=4$}
\setcounter{equation}{0}

In this appendix we briefly summarize the asymptotic behaviour of Maxwell fields ($p=2$) in four dimensions, still assuming a power-like behaviour of the leading terms. As in the main text, it is necessary to consider the cases without/with a cosmological constant separately. Recall that for $p=2$ the case $n=4$ is special essentially due to the term $F_{1j}(\R{ij}+\R{ji})-F_{1i}\R{}=(4-n)F_{1i}^{(\de)}r^{\de-1}+\ldots$ in \eqref{+C1+A2}.

\label{app_4D}

\subsection{Case $R=0$}

If $\a\ge-1$ then 
	\beqn
		& & F_{0i}=F_{0i}^{(\a)}r^\a+\ldots ,  \\
		& & F_{01}=F_{01}^{(\a)}r^\a+\ldots , \qquad F_{ij}=F_{ij}^{(\a)}r^\a+\ldots , \qquad (\a\ge-1) \\
		& & F_{1i}=F_{1i}^{(\a)}r^\a+\ldots . 
	\eeqn

If $-2\le\a<-1$ then 
	\beqn
		& & F_{0i}=F_{0i}^{(\a)}r^\a+\ldots ,  \\
		& & F_{01}=F_{01}^{(\a)}r^\a+\ldots , \qquad F_{ij}=F_{ij}^{(\a)}r^\a+\ldots ,  \qquad (-2\le\a<-1) \\
		& & F_{1i}=\frac{F_{1i}^{(-1)}}{r}+\ldots . 
	\eeqn

Finally, if $\a<-2$ then 
	\beqn
		& & F_{0i}=F_{0i}^{(\a)}r^\a+\ldots ,  \\
		& & F_{01}=\frac{F_{01}^{(-2)}}{r^2}+\ldots , \qquad \frac{F_{ij}^{(-2)}}{r^2}+\ldots ,  \qquad (\a<-2) \\
		& & F_{1i}=\frac{F_{1i}^{(-1)}}{r}+\ldots . 
	\eeqn	
Special cases where only $F_{1i}$ is non-zero (and the field is thus of type N aligned with $\bl$) do exist \cite{RobTra62}.
For $\a=-3$ this fall-off agrees with the standard results \cite{GK,JanNew65,Penrose65prs} (and is also contained in the general result \eqref{+1_n/2}--\eqref{-1_n/2} with $n=4=2p$). Recall, however, the more general analysis of \cite{Kroon00} for cases with $\a>-3$. Some comments for a possible case with $\a>-3$ in the full Einstein-Maxwell theory are also contained in \cite{Penrose65prs}.

\subsection{Case $R\neq0$}

In the presence of a cosmological constant one necessarily has $\a\le -3$ (recall the discussion at the beginning of section~\ref{sec_R}), and the fall-off  

	\beqn
		& & F_{0i}=\frac{F_{0i}^{(-3)}}{r^3}+\ldots ,  \\
		& & F_{01}=\frac{F_{01}^{(-2)}}{r^2}+\ldots , \qquad \frac{F_{ij}^{(-2)}}{r^2}+\ldots ,  \\
		& & F_{1i}=\frac{F_{1i}^{(-1)}}{r}+\ldots , 
	\eeqn	
also in agreement with \cite{Penrose65prs} (and again contained in \eqref{+1_n/2}--\eqref{-1_n/2}). We further observe that if $\a<-3$, then necessarily $\a\le-4$ (because of \eqref{+C1+A2}), but $F_{1i}^{(-1)}$ can still be non-zero, as opposed to the case $n>4$ (section~\ref{sec_R}).


\end{document}